\documentclass{article}

\usepackage{PRIMEarxiv}

\usepackage[utf8]{inputenc} % allow utf-8 input
\usepackage[T1]{fontenc}    % use 8-bit T1 fonts
\usepackage{hyperref}       % hyperlinks
\usepackage{url}            % simple URL typesetting
\usepackage{booktabs}       % professional-quality tables
\usepackage{amsfonts}       % blackboard math symbols
\usepackage{nicefrac}       % compact symbols for 1/2, etc.
\usepackage{microtype}      % microtypography
\usepackage{lipsum}
\usepackage{fancyhdr}       % header
\usepackage{graphicx}       % graphics
\usepackage{makecell, multirow}
\usepackage{amsmath}
\graphicspath{{media/}}     % organize your images and other figures under media/ folder

\title{Topic Community based Temporal Expertise  for Question Routing
}

\author{
  Vaibhav Krishna, Vaiva Vasiliauskaite, Nino Antulov-Fantulin  \\
  ETH Zürich \\
  email: \texttt{vaibhavkrishna@ethz.ch},\\ \texttt{vaiva.vasiliauskaite@gess.ethz.ch},\\
  \texttt{nino.antulov@gess.ethz.ch}
}

\begin{document}
\maketitle
\begin{abstract}
Question Routing in Community-based Question Answering websites aims at recommending newly posted questions to potential users who are most likely to provide ``accepted answers". Most of the existing approaches predict users' expertise based on their past question answering behavior and the content of new questions. However, these approaches suffer from challenges in three aspects: 1) sparsity of users' past records results in lack of personalized recommendation that at times does not match users' interest or domain expertise, 2) modeling based on all questions and answers content makes periodic updates computationally expensive, and 3) while CQA sites are highly dynamic, they are mostly considered as static. This paper proposes a novel approach to QR that addresses the above challenges. It is based on dynamic modeling of users’ activity on topic communities. Experimental results on three real-world datasets demonstrate that the proposed model significantly outperforms competitive baseline models.
\end{abstract}
% keywords can be removed

\section{Introduction}\label{sec1}

The advent of Web 2.0 has led to the popularity of systems that are based on user-generated content \cite{srba2016comprehensive}. One such web-based service that relies on users generating content themselves is known as Community Question Answering (CQA). CQA websites such as \href{http://answers.yahoo.com}{Yahoo! Answers}, \href{https://www.quora.com}{Quora},  \href{https://stackoverflow.com}{Stack Overflow} and \href{https://wiki-answers.com}{Wikianswers} are becoming increasingly important for sharing and spreading knowledge \cite{srba2016comprehensive,yuan2020expert}. These platforms leverage ``the wisdom of crowds"\cite{surowiecki2005wisdom} and provide a venue where multiple users can exchange information in the form of questions and answers \cite{yang2015wise}. While CQA services offer significant help to knowledge seekers, their rapid growth poses unique challenges. First, these sites witness thousands of questions posted every day in addition to millions of questions that already exist. This makes it difficult for an answerer to find the appropriate question that matches his or her expertise \cite{chang2013routing}. Second, as the expertise and education levels vary a lot among answerers, the quality of answers received is difficult to control. Third, the increasing time to receive a high-quality answer and a growing number of low-quality answers cause a high \emph{churn rate} (the rate at which users leave the community or become inactive), hampering the sustainability of these CQA systems \cite{srba2016stack}.

Past studies have shown that the CQA communities have only a handful of domain experts that provide most of the high-quality answers \cite{sung2013booming}. Acknowledging this long-tail distribution of users' answering quality, one could resolve the above challenges by automatically identifying the ``experts" that tend to contribute high-quality answers and direct unanswered questions to them. An algorithm that successfully links users to questions they are likely to provide good answers to is known as question routing (QR) \cite{li2011question}. The success of QR can potentially increase the participation rates of users and foster stronger communities in CQA. Approaches to question routing make use of various data science tools such as information retrieval (IR), machine learning, natural language processing (NLP), and social computing perspective \cite{wang2018survey,al2018understanding}. 

However, despite the active research in CQA, QR remains a challenging task which can be attributed to three key limitations. First, the sparsity of users' historical question and answer records makes it difficult to infer their domain expertise \cite{le2016retrieving}. This results in a lack of personalized recommendations. Second, most of the CQA systems are dynamic: new users join constantly, and some accounts become abandoned. As a result, evaluating available experts for a newly posted question would require periodic updates. Thus, running complex models that use all possible data is becoming increasingly costly. Third, most of the past studies have assumed CQA sites as static environments, while overlooking the temporal aspects such as users' willingness to contribute, changing interests, and knowledge evolution \cite{pal2012evolution}. 
These characteristics indicate the need for self-evolving approaches for expert recommendation that can be updated efficiently when new information is available.

\begin{figure}
\includegraphics[scale=0.6]{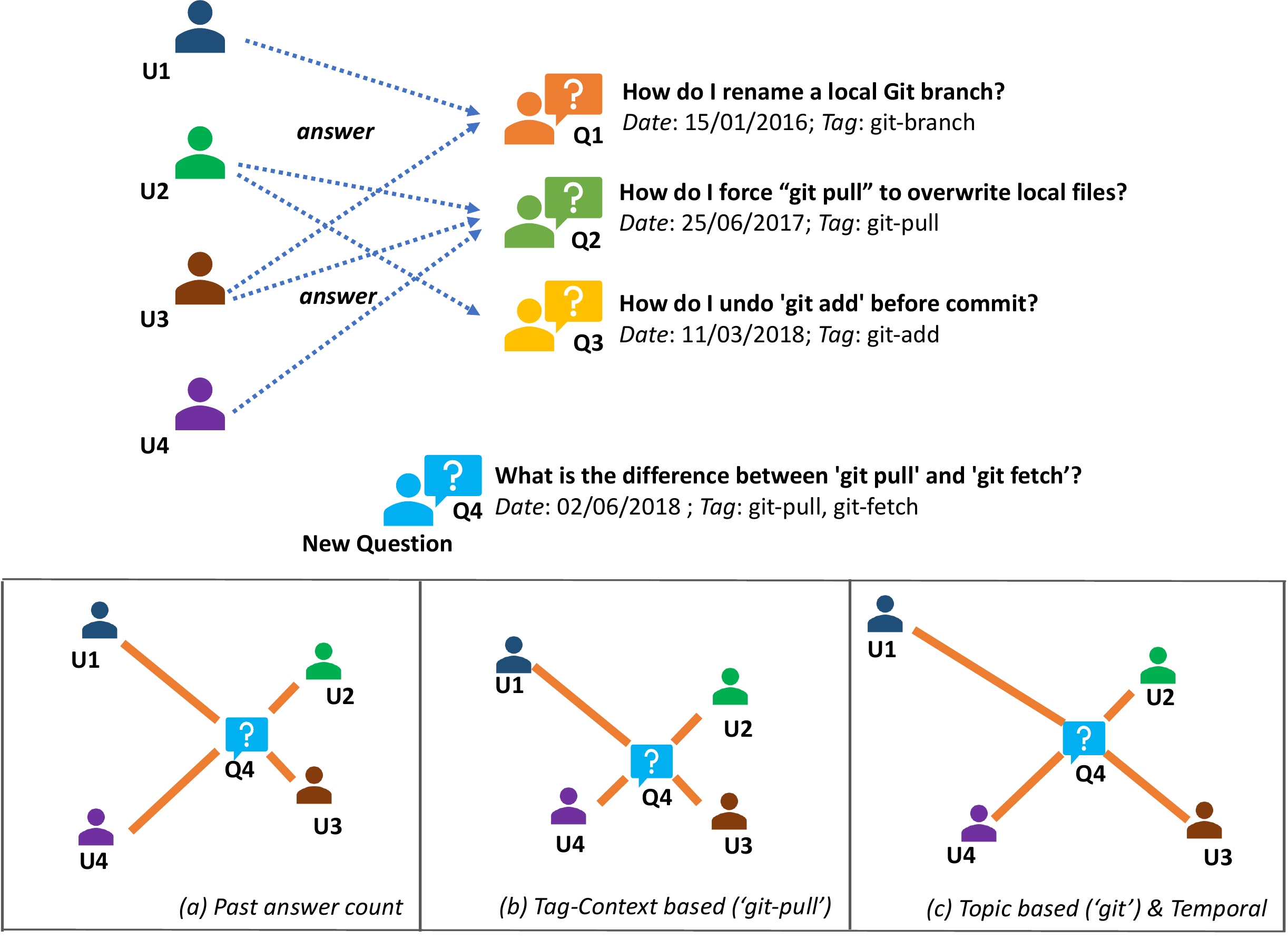}
\caption{{\bf {An example of question-routing based on different types of metadata.}} \\
(a) considering previous answer count, answerers U2 and U3 have equal expertise for a new question Q4 (both having shortest path); (b) considering context of the previous questions based on tags, Q4 is closest to Q2 (tag 'git-pull' in this case). Thus, answerers U2, U3 and U4 be considered having equal expertise (shortest path for all three users). (c) considering topic community - tags in all the questions Q1, Q2, Q3 and Q4 are correlated to topic 'git'. Thus, answerers U2 and U3 be ranked higher. However, considering the temporal aspect, U2 is more likely to answer question Q4 (only shortest path).
\label{fig1}}
\end{figure}

To overcome these limitations, we propose a novel topic community based temporal expertise for question routing (TCTE-QR). 
Our contribution in this work is three-fold and can be summarised as follows:

1)	We introduced a novel personalized recommendation method that uses a key feature of CQA platforms - the semantic similarity networks of questions are modular. The proposed recommender framework is designed to consider the  similarity between the content of the recent activity of a user (the ``domain expertise") and the content of a given new question. To find the domain expertise of a user we use the topics of the archived questions answered by the user.

2) Compared to past studies that relied on all available textual data to infer topic experts, in our approach we use only tags assigned to questions. We define a ``tag graph'' using association rule mining \cite{agrawal1993mining} and subsequently used community detection to infer topics from the tag graph and topic experts from users' past activities on these topics. In this way, the proposed method can learn domain experts efficiently with reduced computational cost, facilitating routine implementation.

3) Another key feature that TCTE-QR leverages is that users' interests and expertise change over time. Thus, the knowledge of a user is determined using the questions recently answered by the user. To incorporate the evolving interest of the user, we use a decay function to the archived answers, giving more weight to recent answering activities. 
 
The rest of this paper is organised as follows. In Section 2, we briefly review the standard approaches of question routing. Next, in Section 3, we look at relevant properties of CQA platforms and discuss the task of question routing. In Section 4, we present our proposed approach --- ``Topic Community-based Temporal Expertise Question Routing (TCTE-QR)''. In Section 5, we describe the models and the data used to test our algorithm. In Section 6, we evaluate the proposed model to several baseline methods, summarizing the article in Section 7.

\section{Related Work}\label{sec2}

In the past few years, question routing has attracted a lot of attention in the IR community \cite{neshati2017dynamicity}. We briefly review prior research on question routing and classify the approaches into the following four broad categories: classification-based, network-based, text-based, and collaborative filtering methods. Further, we highlight the major challenges in each of these approaches.

\subsection{Classification-based approaches}
Question routing can be translated to a problem of identifying experts as a class of users among all users and recommending questions to them. Thus this can be solved as a classification problem that aims to distinguish the expert class of users from others. The advantage of classification methods is that these methods can easily apply multiple features of users, questions, answers, or interaction networks. For example, Pal and Konstan \cite{pal2010expert} used question features (e.g., question length), user features (e.g., the previous number of best answers), and user feedback on answers, training a binary classifier to distinguish experts from others.
Similarly, Zhou et al. \cite{zhou2012classification} built a classifier using local and global features of questions, user history, and question-user relationships. Commonly used classifiers in the above approaches are Support Vector Machine \cite{ji2013learning}, Random Forest \cite{choetkiertikul2015will}, and Naive Bayes \cite{van2015early}.
However, these approaches are limited, as hand-crafted feature extraction is required. It is not only time-consuming but also reliant on selected features that can cause selection bias. 

\subsection{Network-based approaches}
Network-based approaches analyze a user-user network formed by their asking-answering relationships. Next, a link analysis technique \cite{borodin2005link} is used on the network to evaluate the authority of each user. The simplest way to measure the authority of a user in the CQA community is by using degree centrality measure --- InDegree \cite{zhang2007expertise}, which considers users who have answered more questions in the user-user network as better answerers \cite{jeon2006framework}. Another notable approach is the community expertise network (CEN) that uses ``z-score" to measure the authority of an answerer based on in-degrees and out-degrees \cite{zhang2007expertise}. Other centrality measures are also used, e.g., ExpertiseRank, a slight variant of PageRank \cite{page1999pagerank}, or HITS \cite{kleinberg1999authoritative}. 
A major challenge of methods based on centrality measures is that they cannot leverage textual information, topics, categories of questions. Thus most of these approaches recommend questions to users based on their general expertise (rather than their expertise on particular topics).

\subsection{Text-based approaches}
Another set of approaches builds recommendations around topics inferred using language and topic models from the text of questions and answers  \cite{li2011question}. 
The language models use a generative approach to compute the word-based relevance of a user's past activities to a new question and compute the probability of the user answering the question \cite{zheng2012algorithm}. Finally, a ranked list of users based on their likelihood of answering the given question is generated. However, language models are based on exact word matching, therefore they are not able to capture advanced semantics (``lexical gap") \cite{zhou2012topic}. In addition, data sparseness and co-occurrence of irrelevant words in user profiles or questions can lead to word mismatch between the routed question and user profiles.

To bypass the lexical gap, topic models were introduced that measure relationships in the topic space rather than the word space and thus do not require the exact word to appear in the user profile \cite{riahi2012finding}. One of the most widely used topic models is Latent Dirichlet allocation (LDA) in which topic mixture is drawn from a conjugate Dirichlet prior that remains the same for all users \cite{blei2003latent}. In such QR approaches, LDA is first used to extract topics based on users' past activity that shows the connection between the expert users and new questions. In the second step, these topics are used to compute the probability of each user to provide an answer, ordering users based on this probability \cite{momtazi2013topic, zhou2012topic}. However, a limitation of LDA based method is that standard LDA groups all users' questions under one topic. Riahi et al.\cite{riahi2012finding} proposed segmented topic model to overcome this limitation, which allows questions to have different topical distributions. 

In addition, there are other studies based on a hybrid approach that leverage both topic relevance as well user authority for expert finding in CQA. For example, Kao et al.\cite{kao2010expert} used user reputation and category into link analysis for expert finding, while in related studies topic-sensitive probabilistic model was combined with PageRank to improve the recommendation methods based on the latter (see \cite{li2015hybrid,zhao2014expert}). Yang et al.\cite{yang2013cqarank} used topic modeling in Topic Expertise Model (TEM) and combined it with link analysis among users to recommend experts.  

A major challenge of these approaches is that they require all of the questions and answers data to extract topics. This makes them computationally expensive in dynamic environments like CQA services where periodic updates are useful. Further, since probabilistic models such as LDA distribute the total probability of 1 among all the topics for each user, having a higher probability on one topic for a user will discount his probability on other topics \cite{wang2018survey}. However, a user can have great expertise in multiple topics simultaneously. 

\subsection{Collaborative filtering approaches}
Another stream of research applied collaborative filtering (CF) methods such as matrix factorization (MF) techniques which are known to be advantageous in terms of flexibility and scalability in the recommendation domain \cite{koren2009matrix}. For example, Zhao et al. \cite{zhao2014expert} used MF by representing the questions with their content words, defined a user-word matrix to discover the user's expertise on particular words. However, this results in a high-dimensional, sparse matrix which affects the performance of the MF approach \cite{wang2016personalized,idrissi2020systematic}. Further, as items in the matrix factorization approach are treated as independent, the semantic similarity between words is ignored. 

Advancing in this direction, Yang et al. \cite{yang2014tag} proposed to use \emph{tags} (keywords), that summarize the question's focus, instead of the textual content of questions and answers to perform the MF for QR. The study highlighted that tags are more informative and well-summarising the topic of focus of the question. The approach also uses the number of votes for an answer to evaluate the answerer's expertise in a given question and related tags. This method outperformed other, text-based topic modeling approaches like TEM \cite{yang2013cqarank}, and was shown to be several orders of magnitude faster. In particular, MF learns the latent feature space of both users and tags (words) to build a user-tag (user-word) matrix which is used to recommend experts given a new question. 

\begin{figure}[!h]
\includegraphics[scale=0.5]{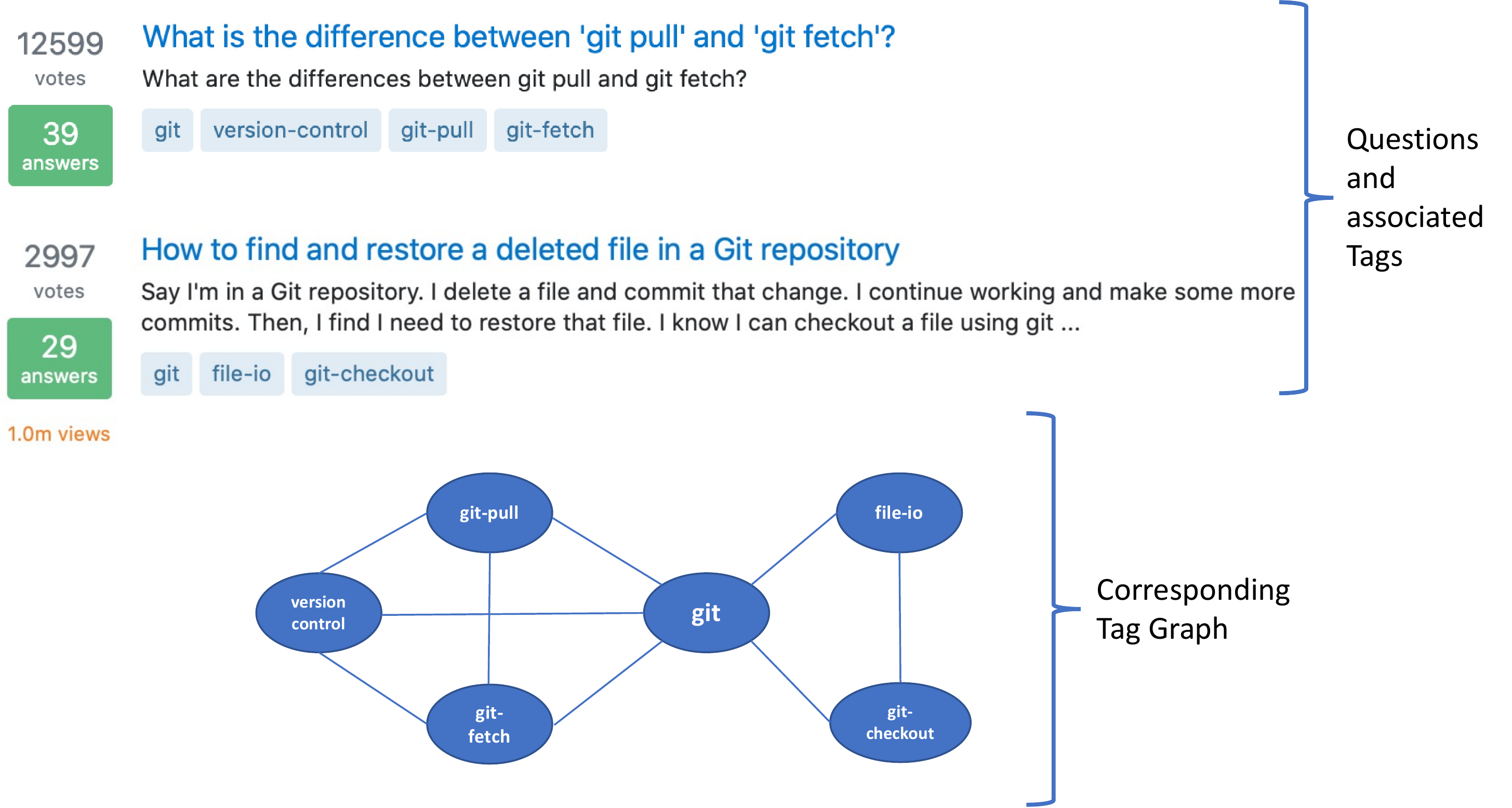}
\caption{{Question post from StackExchange CQA network} \label{fig_question_sample}}
\end{figure}

However, while using tags instead of words (i.e., textual content of questions and answers) overcomes the dimensionality issue to some extent, it does not completely resolve the problem of data sparsity. In addition to that, the approach still suffers from having a high number of correlated items. Both of these issues affect MF performance \cite{najafabadi2016systematic,huang2004applying,idrissi2020systematic}. For example, while tags such as “java”, “java-9”, “jdk” are arguably similar, they are treated as independent when applying matrix factorization based on user-tag matrix. Figure \ref{fig_question_sample} shows an example of two topically similar questions and their seemingly similar tags. A recent study by Fukui et al. \cite{fukui2019suggesting} tried to improve the approach of \cite{yang2014tag} by expanding tagged keywords based on word embeddings to mitigate one of the issues related to the spelling variants of tagged keywords such as “java” and “java-9”. 

In addition to the issues highlighted above, most of the literature considers the expertise in a static environment (i.e.\ at the new question time), ignoring the evolution of personal expertise and interest over time \cite{neshati2017dynamicity}. 

In summary, although many innovative approaches to question routing have been proposed over the years, there are several outstanding issues and possible areas of improvement which we aim to address in this work. 

\section{Problem formalization}\label{sec3}

\subsection{Basic properties of community question-answer platforms}

In this section, we define the CQA platform as a network. Topologically, the platform consists of a \emph{global sample} of questions, users, and tags. First, we have a set of questions $ \mathcal{Q} = \{q_i\} $, $\vert \mathcal{Q} \vert=Q$, indexed over integers $i,j\in [1,Q]$ and a set of $M$ answerers (users that may respond to questions) $\mathcal{U}= \{u_{\alpha},u_{\beta},...\}$, indexed over $\alpha,\beta$, and the size of this set $\vert \mathcal{U} \vert =M$. Lastly, we have tags $\tau$ that form a set $\mathcal{V}$; tags are indexed with integers $\kappa$ that range from $1$ to $K$. Each question is associated with a tag if the tag features in the question. We then write $\tau \sim q$. Each user is also associated with questions and their related tags: we will write that $u \sim q$ if a question $q$ is addressed by $u$. By transitivity, if $q \sim \tau$, then $u \sim  \tau$.
 
Each question $q_i\in  \mathcal{Q} $ is mapped to a tuple $q_i=(\mathcal{C}^{(i)}, t_i)$ where $\mathcal{C}^{(i)}$ is some set describing the question's contents. For example, $\mathcal{C}^{(i)}$ can be obtained using natural language processing, topic modeling, tags, etc. In tag-based QR approach, tags of each question are used to describe $\mathcal{C}^{(i)}$. $t_i$ is the timestamp when the question was posed. 

Each \emph{question-answer session} comprises of a question $q_i$ and a subset of answerers $\mathcal{U}_i$ that responded to the question $q_i$:
\begin{equation}
    \mathcal{U}_i = \{ u \in \mathcal{U} \quad \forall \quad u\sim q_i\},
\end{equation}
We further denote the user who gave the accepted answer by $u^{*},u^{*}\in \mathcal{U}_i$, i.e., the user whose answer is considered as fully addressing the question $q_i$.

\subsection{Question routing}

Question routing is the task of predicting a user in CQA who is most likely to share knowledge and answer a newly posted question~\cite{zhou2012classification}. In general, the \emph{Question Routing} task ranks a users' set $\mathcal{U}$ to obtain an ordered set $\mathcal{U}^{\textrm{rank}}$ such that the order reflects on how likely a user will provide an answer to a new question $\hat{q}=(\hat{\mathcal{C}},\hat{t})\notin \mathcal{Q}$: the higher in the ranking $u$ is, the more likely the user is to answer $\hat{q}$. We essentially endow the set $\mathcal{U}$ with order relations ``$\prec$", e.g., $u_{\beta}\prec u_{\alpha} \in \mathcal{U}^{\textrm{rank}}$ if we find that user $u_{\alpha}$ is more likely to respond to $\hat{q}$ than a user $u_{\beta}$. In reality, the set $\mathcal{U}^{\textrm{rank}}$ need not be totally ordered (i.e., we may find more than one user that is equally likely to respond to $\hat{q}$), yielding a partially ordered set $\mathcal{U}^{\textrm{rank}}$, but in practice, this is very unlikely, as we will essentially be mapping the users to real-valued numbers to produce this order. Finally, an order relation $\prec$ reflects on the deduced rank relation eq-\ref{eq_order}:

\begin{equation}\label{eq_order}
  \mathcal{U}^{\textrm{rank}}= \begin{cases}   u_{\beta} \succ u_{\alpha} \in  \mathcal{U}   \quad  \mid \quad \textrm{rank}(u_{\beta},\hat{q}) >\textrm{rank}(u_{\alpha},\hat{q}) \\
  u_{\beta} \prec u_{\alpha} \in\mathcal{U} \quad  \mid \quad \textrm{rank}(u_{\beta},\hat{q}) <\textrm{rank}(u_{\alpha},\hat{q})
  \end{cases}.
\end{equation}

We also define $u^*$ as the supremum of $\mathcal{U}^{\textrm{rank}}$:
\begin{equation}
    u^* = \sup (\mathcal{U}^{\textrm{rank}}). 
\end{equation}

\section{Proposed Model - TCTE-QR}\label{sec_tcteqr}

Our approach consists of four steps:

\paragraph{Step 1: Building topic communities given a projection of a bipartite network of questions and tags}\label{step_1}

As shown in~\cite{yang2014tag}, the expertise of the user answering a question can be viewed as his/her expertise on the tags of this question. In our approach, we propose using a collection of similar tags $\mathcal{C}_{\omega} =\{\tau\}$, as opposed to each individual tag for quantifying the user's activity/expertise within the topic $\mathcal{C}_{\omega}$ the tag represents.

To systematically obtain collections of similar tags $\{\tau_{\kappa}\}$, we consider a weighted undirected graph $\mathcal{G}=(\mathcal{V},\mathcal{E},\mathcal{W})$, where the node-set $\mathcal{V}$ contains the tags $\tau_{\kappa}$, the edge set $\mathcal{E}$ contains undirected edges $(\tau_{\kappa}, \tau_{\lambda})$, and the weights set $\mathcal{W}$ contains information about the edge weights. We define the weight of an edge $(\tau_{\kappa}, \tau_{\lambda})$, $w_{\lambda\kappa}$ as the number of questions associated with a given pair of tags. An edge set is defined as:
\begin{equation}
   \mathcal{E} = \{ (\tau_{\kappa}, \tau_{\lambda}) ,\tau_{\kappa}, \tau_{\lambda}\in \mathcal{V} \mbox{ if } \exists  q_i \in\mathcal{Q} \quad \mid \quad q_i \sim \tau_{\kappa}, q_i\sim \tau_{\lambda}\},
    \label{eq_tag_edge}
\end{equation}
and for each edge its weight is defined as:
\begin{eqnarray}
    w_{\lambda\kappa} = \mid \{q_i \sim \tau_{\kappa} \textrm{ and } q_i\sim \tau_{\lambda} \mbox{ for } q_i \in \mathcal{Q}\} \mid 
\end{eqnarray}

Further, we consider the non-unitary minimum value of questions that each pair of tags needs to be collectively associated with, $N_q$, in order to confirm the edge in $\mathcal{G}$. Thus, we have a projection of a bipartite network between questions and tags, projected onto a tag layer.

We expect $\mathcal{G}$ to be modular: some relevant tags would be strongly connected to each other, but loosely connected to irrelevant tags. In light of this, we apply community detection that maximises partition quality given by eq-\ref{modular} to separate tags into communities $\mathcal{C}$. We use these communities as a proxy for topics in the consequent steps. 

\begin{equation}
Q(\mathcal{G},\mathcal{C}) =  \frac{1}{2m} \sum_{\kappa,\lambda} \left(A_{\kappa\lambda} - \frac{d_\kappa d_\lambda}{2m}\right)\phi(c_{\kappa}, c_{\lambda})
\label{modular}
\end{equation} 

where for the tag graph $\mathcal{G}$ and given partition $\mathcal{C}$, $A$ is the adjacency matrix, $m$ is the total number of edges in $\mathcal{G}$, $d_\kappa$ is the degree of the node $\tau_\kappa$ and function $\phi(c_{\kappa}, c_{\lambda})$ is $1$ if tags $(\tau_{\kappa}, \tau_{\lambda})$ are in the same community and $0$ otherwise.

\paragraph{Step 2: Creating temporal user-topic activity matrix $\mathbf{S}(\hat{t})$ that accounts for user's activities within topics, and for temporal changes in users' response patterns}

Once the tags are clustered into topic communities as described above, a \emph{user-topic activity matrix} is created. For each user, all positively scored answers and the respective question along with associated tags are collected. Next, the tags are mapped to the topic communities discovered in the previous step. Accordingly, the user is assigned an activity score on a topic the question is related to. This score is calculated as the fraction of tags associated with the question that comes from the topic out of all the tags, related to the question:
\begin{equation}
    f_{q_i}^{\alpha\omega} =\frac{\mid \{\tau \in \mathcal{C}_{\omega} \mid u_\alpha\sim\tau\sim q_i\} \mid}{\mid \{\tau \in \mathcal{V} \mid u_\alpha\sim\tau\sim q_i\} \mid}.
\end{equation}
The net activity score of the user for a given topic is the sum of all the activity-score on all the questions related to that topic:

\begin{equation}
s_{\alpha \omega} = \sum_{q_i \in\mathcal{Q}}f_{q_i}^{\alpha\omega}
\label{eq_sim_t}
\end{equation}

\paragraph{Temporal discounting}
\emph{Temporal discounting}~\cite{green1994temporal} is a term that refers to the tendency of people to give more value to near-future rewards while discounting delayed rewards. Similarly, in a dynamic environment where the users’ activities keep changing, the system should give more value to users with recent activities and discount earlier activities.

Consider the CQA site where the first post occurred at $t_1=1$ and a new question $\hat{q}$ is posted at time $t_q=\hat{t}$. To identify the expert candidates who can reply to question $\hat{q}$ all the positively scored answers by the users posted within the period $[t_1,t_q]$ are considered. Next, the time period $[t_1,t_q]$ is divided into time windows $\delta$ such as day, week, or month. The answers are grouped according to their corresponding time window determined by the date of the answers. Next, a user-topic activity matrix $\mathbf{S}_j$ is defined, where an entry $s_{\alpha \omega}^{j}$ corresponds to the user $u_\alpha$ activity on topic $\omega$ for the time-window $[t_{j-1},t_{j}]$: 

\begin{equation}
s_{\alpha \omega}^{j} = \sum_{q_i \in\mathcal{Q}_{[t_{j-1},t_{j}]}}f_{q_i}^{\alpha\omega}.
\label{temporal_activity}
\end{equation} 

To account for the temporal changes in users' activities, we consider a temporal kernel (discounting function) of a form

\begin{equation}
\label{eq_hyper_disc}
    g(j)=\frac{1}{1+j},
\end{equation} 
where $j$ is the number of time windows $\delta$ that passed from the time of interest until present.

\begin{figure}
\includegraphics[scale=0.25]{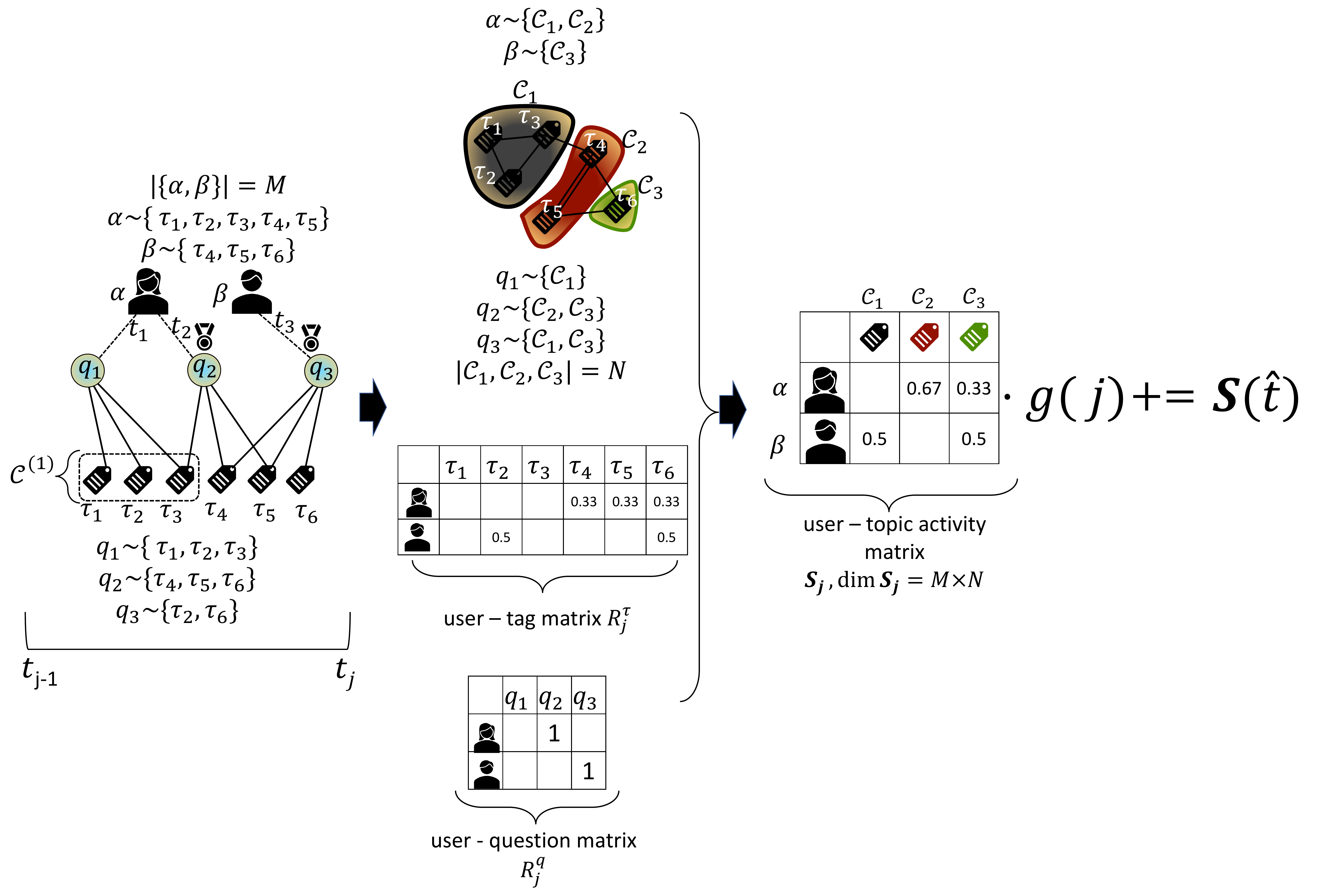}
\caption{Illustration of workflow to obtain the temporal user-topic activity matrix $\textbf{S}(\hat{t})$ for a given time window $[t_{j-1},t_j]$ and the current time $\hat{t}$.\\
The first step is to find communities of tags, which are then assumed to be topics $\mathcal{C}_{\omega}$. We combine the information about all the positively scored answers on questions answered by users within the time window into $\textbf{R}^{q}_j$. Further, to avoid double counting, the entries for user-tag matrix is estimated based on the proportion of tags. Thus, for a question $q_2$ with three tags, each of the tags gets an entry = 0.33 in $\textbf{R}^{\tau}_{j}$. We then combine this and information about which topic communities the questions belong to to obtain each entry $s_{\alpha\omega}$. In particular, for a user $u_\alpha$ a contribution towards the black topic $\mathcal{C}_1$ is not accounted for, because the user did not receive a positive score for the answer on question $q_1$, the only associated with this topic community. Therefore $u_\alpha$ has $s_{\alpha\omega}=0.67$ for a red topic community and $s_{\alpha\omega}=0.33$ for a green topic community. This contribution comes from the positively scored answer to the question $q_2$. \\The obtained matrix $\textbf{S}_j$ is weighted by a function $g(j)$ and the result is added to the temporally discounted user activity matrix $\textbf{S}(\hat{t})$.
\label{fig_temp_expertise}}

\end{figure}

Finally, given this temporal division of users answering activity, we define a temporal user-topic activity matrix $\mathbf{S}(\hat{t})$, where an entry $s_{\alpha \omega}^{\hat{t}}$ corresponds to the user $u_\alpha$ temporal-discounted activity on topic $\omega$.

\begin{equation}
\mathbf{S}({\hat{t}}) = \sum_{j=1}^{J}g(j) \mathbf{S}_{j},
\label{temporal_activity_Netscr}
\end{equation} 
where, ${J}$ is the total number of time windows.

We illustrate the process of clustering topics into communities and building $\mathbf{S}(\hat{t})$ in Figure \ref{fig_temp_expertise}. This matrix is of size $M \times N $ where $N$ is the number topics and $M$ is the number of users.

\paragraph{Step 3: Factorising the user-topic activity matrix to obtain the user-topic \emph{expertise} matrix $\textbf{U}\textbf{T}^\top$}

\begin{figure}
    \centering
    \includegraphics[scale=0.2]{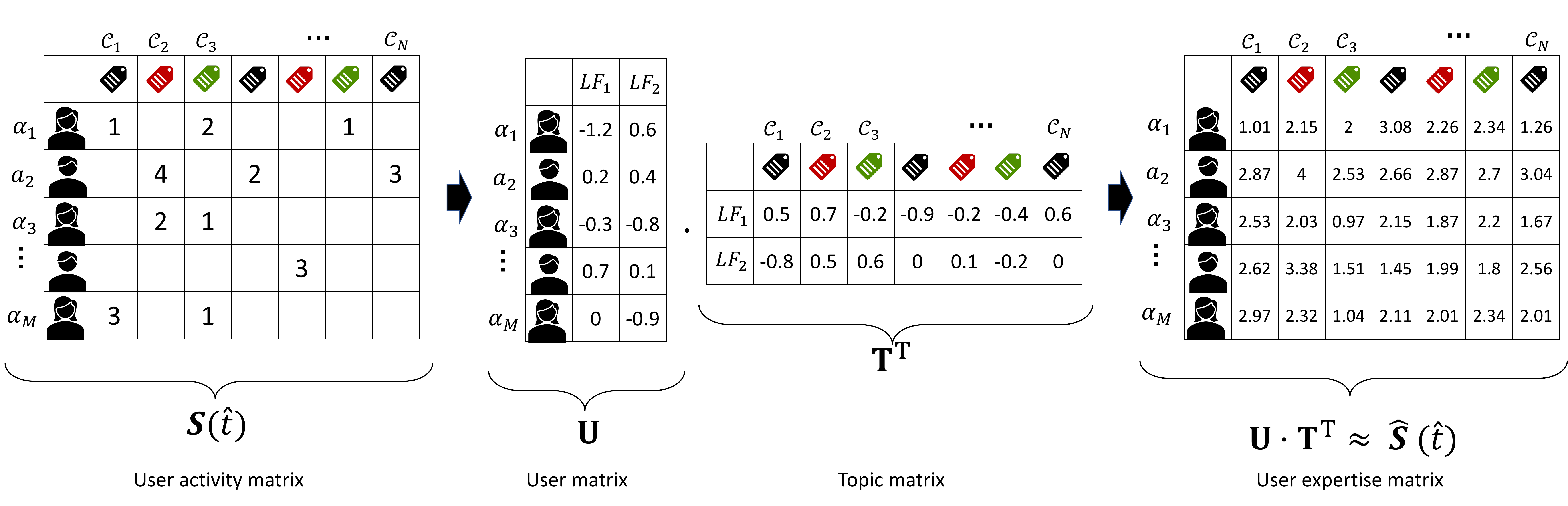}
    \caption{Illustration of a workflow of matrix factorization. \label{fig_mf}}
    
\end{figure}

We perform matrix factorization \cite{koren2009matrix} on the user-topic activity matrix, to learn the latent features of users and topics. Since not all users answered questions from all topics, $\mathbf{S}(\hat{t})$ is \emph{incomplete}. Therefore, using matrix factorization, the goal is to find an approximation of $\mathbf{S}(\hat{t})$. In practice, this is done by mapping users and  items (topics) to a joint latent-factor space of dimensionality $\ell$ lower than $M$ for users and $N$ for topics.
The mapping is done such that the inner product of the users feature matrix and topics feature matrix approximates users' expertise on topics. This matrix factorization is illustrated in Figure \ref{fig_mf}.

The algorithm begins by creating a users' feature matrix $\mathbf{U}$ that is of size $M\times \ell$, and a topic feature matrix $\mathbf{T}$ of size $N \times \ell$. Here $\ell$ is the number of latent features.
Initially, the entries of each matrix are assigned at random, sampling from a Gaussian distribution. Thus, each user $u_{\alpha}$, and topic $\mathcal{C}_{\omega}$ are associated with vectors $\mathbf{u}_\alpha \in\mathbb{R}^{\ell}$ and $\mathbf{t}_{\omega} \in\mathbb{R}^{\ell} $ respectively. A dot product $\mathbf{u}_\alpha \mathbf{t}_{\omega}^{\top}$ captures the overall interest of the user $u_{\alpha}$ in the topic $\mathcal{C}_{\omega}$ and approximates user's expertise-score in the topic $s_{\alpha \omega}^{\hat{t}}$. 

In order to learn all entries of the vectors $\mathbf{u}_\alpha$ and $\mathbf{t}_{\omega}$, namely the ``latent features", the model minimises the regularised squared error on the set of known activity score:

\begin{equation}\label{eq_matrix_fac}
\varepsilon = \sum_{(\alpha,\omega) \mid \mathcal{C}_{\omega} \sim u_{\alpha}}\varepsilon_{\alpha\omega} = \min_{u_* \mathcal{C}_*} \sum_{(\alpha,\omega) \mid \mathcal{C}_{\omega} \sim u_{\alpha}} \left(s_{\alpha \omega}^{\hat{t}} -  \mathbf{u}_\alpha \mathbf{t}_{\omega}^{\top} \right)^{2} + \lambda ( \| \mathbf{u}_\alpha  \|_{\mathcal{F}}^2 + \| \mathbf{t}_{\omega}  \|_{\mathcal{F}}^2),
\end{equation}

where,
$(\alpha,\omega) \mid \mathcal{C}_{\omega} \sim u_{\alpha}$ is the set of user-topic index pairs for which $s_{\alpha \omega}^{\hat{t}}$ is known, and
$\| \cdot \|_{\mathcal{F}}^2$ is a Frobenius norm (also known as L2-norm).

The constant $\lambda$ controls the extent of regularization and avoids overfitting the observed data. The terms in the right-most bracket are known as Tikhonov's terms. Commonly used approaches to minimize eq \ref{eq_matrix_fac} are stochastic gradient descent and alternating least squares~\cite{koren2009matrix}. The optimal values for the hyperparameters such as the number of latent factors $\ell$, the regularization parameter $\lambda$, and the learning rate are determined by k-fold cross validation on the training dataset \cite{bishop2006pattern}.

\paragraph{Step 4: Expert Recommendation for a new question}

For a newly posted question $\hat{q}$, first the tags of the question are mapped to topic communities and the score is calculated for each user $u_\alpha$:
\begin{equation}\label{eq_rank}
\textrm{rank}(u_\alpha,\hat{q}) = \sum_{\mathcal{C}_{\omega} \mid \exists \tau\in \mathcal{C}_{\omega},\ \hat{q}\sim \tau } w_\omega \left[\mathbf{U} \mathbf{T}^{\top}\right]_{\alpha\omega},
\end{equation}
where  $w_\omega$ is a weight---importance of the topic community---in the context of $\hat{q}$, calculated as the fraction of tags associated with the question that come from the community $\mathcal{C}_{\omega}$ out of all tags, related to the question:
\begin{equation}
    w_\omega =\frac{\mid \{\tau \in \mathcal{C}_{\omega} \mid \tau\sim \hat{q}\} \mid}{\mid\{\tau \in \mathcal{V} \mid \tau\sim \hat{q}\} \mid}=\frac{\mid \mathcal{C}_{\omega}\cap \hat{\mathcal{C}}\mid}{\mid\hat{\mathcal{C}}\mid}.
\end{equation}

\begin{figure}
    \centering
    \includegraphics[scale=0.3]{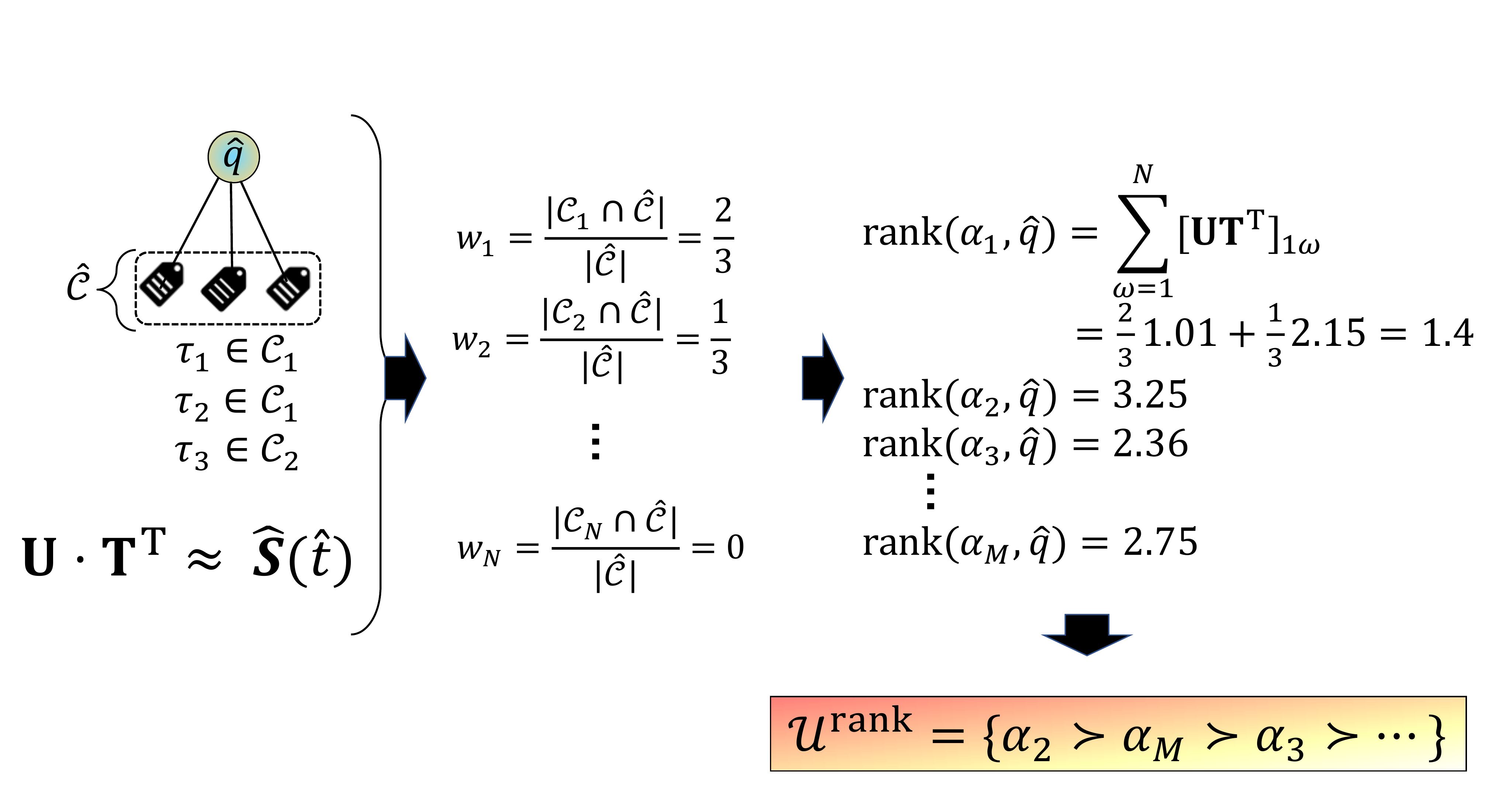}
    \caption{Step 4 of the approach, following the example of user activity matrix deduced via matrix factorization as exemplified in Figure \ref{fig_mf}.}
    \label{fig_recommendation}
\end{figure}

The ordered set $\mathcal{U}^{\textrm{rank}}$ is obtained using the rank of eq-\ref{eq_rank}, as per order relation eq-\ref{eq_order}. The last step of the process is illustrated in Figure \ref{fig_recommendation}.

\section{Data and models}\label{sec5}

\subsection{Experimental setup}\label{expt_setup}

In this section, the performance of the proposed model is evaluated using three datasets. Here we introduce the experiment settings: datasets, baseline models used to evaluate the performance of our method, and different metrics for comparing the baselines with the obtained results. 

\paragraph{Datasets} We considered three CQA datasets from StackExchange (SE) network to evaluate the performance of our model. All data is available online on the \href{https://archive.org/download/stackexchange}{archive} which contains a full history of every SE community, including all the digital footprint of users activity: time-stamped posts (questions, answers), votes on questions and answers, and tags used.

We used data from the three CQAs, namely: \emph{Super User}, \emph{Server Fault}, and \emph{Ask Ubuntu}. The data for each of these three CQA platforms were retrieved from the archive platform that contains all the data from inception until $31^{\textrm{st}}$ December 2020. From each site's archive, we downloaded the XML files, namely \textsc{Posts.xml} and \textsc{Tags.xml}. The \textsc{Tags.xml} file includes the tag information and the date on which each tag was created. \textsc{Posts.xml} file includes all information on timestamped questions and answers, corresponding votes, and the users who made that post. The questions also include the annotated tags and the corresponding accepted answer. For implementing our proposed approach we used all the data from $1^{\textrm{st}}$ January 2015 until $31^{\textrm{st}}$ December 2018 as training data. Further, data from $1^{\textrm{st}}$ January 2019 to $31^{\textrm{st}}$ March 2019 is used to evaluate the performance of our model. The details of the training datasets are presented in Table \ref{table1}, including the link to download each dataset.

\begin{table}[!ht]
\caption{Summary statistics for three datasets considered in the main text. \label{table1}}
\centering
\begin{tabular}{|p{2cm}|p{3cm}|l|l|l|l|}
\hline
Site & Content & Questions  & Answers & Answerers & Tags \\ \hline
\href{https://archive.org/download/stackexchange/superuser.com.7z}{Super User}  & Computer enthusiasts and power users & 145368  & 219795 & 95849 & 4754  \\ \hline
\href{https://archive.org/download/stackexchange/askubuntu.com.7z}{Ask Ubuntu}
& Ubuntu users and developers & 147323 & 193323 & 85608 & 2849  \\ \hline
\href{https://archive.org/download/stackexchange/serverfault.com.7z}{Server Fault} & System and network administrators & 83174 & 136395 & 56871 & 3359 \\ \hline
\end{tabular}
\end{table}

\paragraph{Topic communities} The first part of building the temporal expertise matrix is obtaining topics. For this, we considered tag network $\mathcal{G}$, obtained from the training dataset. Note that in StackExchange, a user can tag a question with a maximum of five tags from a predefined list. We used \textsc{Tags.xml} file that contains tags and their creation date to analyze the temporal nature of the tag list. We found that 90\% of the tags were created on and before $1^{\textrm{st}}$ January 2015 for all datasets. Considering that a list of tags has remained relatively constant for most of the datasets, we analyzed the tag-tag network and topic communities therein without temporal changes.

We started by creating the tag graph using tag pairs that co-occurred in the questions, as discussed in Sec \ref{step_1} (Step-1). Next, for each tag pair, we calculated the number of questions associated with the pair in the dataset and defined a threshold value $N_q$, setting it to be greater than or equal to 5. Next, we used the Louvain method \cite{blondel2008fast} to find topic communities.

\paragraph{Testing and training data} To perform matrix factorization that is at the heart of the proposed approach, we used training data to form $\textbf{S}(\hat{t})$. Following the settings in the previous studies~\cite{li2019personalized}, we filtered all users who provided less than five answers in the training data set to avoid the \emph{cold start problem} \cite{najafabadi2016systematic}. The elements in the user-topic activity matrix were calculated using hyperbolic temporal discounting function $g(j)$, as per eq-\ref{temporal_activity_Netscr}. The time period was divided into time windows of size $\delta$ equal to $1$ month. Further, given the ground truth for a question is the user who has provided the accepted answer, the test set was filtered to have only the questions with an accepted answer.

\paragraph{Quality Metrics} We considered three popular rank evaluation metrics. 

\begin{enumerate}
    \item Mean Reciprocal Rank (MRR): for each of the $Q$ questions in the test dataset:
    \begin{equation}
        \textrm{MRR}=\frac{1}{Q}\sum_{q\in \mathcal{Q}}\frac{1}{ \mid \left\{v\in \mathcal{U}^{\textrm{rank}} \mid v\succ u^*\right\} \mid}.
    \end{equation}
    \item ``Precision@$r$'' (P$@r$): the fraction of questions for which the recommendation provided $u^*$ in one of the $r$ first items of $\mathcal{U}^{\textrm{rank}}$:
    \begin{equation}
        \textrm{P}@r=\frac{1}{Q}\sum_{q\in \mathcal{Q}}\begin{cases}1 \text{ if } \mid \left\{v\in \mathcal{U}^{\textrm{rank}} \mid v\succ u^*\right\} \mid  \leq r\\ 
        0 \text{ otherwise,}
        \end{cases}
    \end{equation}
    We considered $r=5,10$ and reported P@5, P@10.
    
\end{enumerate}

\paragraph{Comparison to other approaches} We compared our method to four alternatives:
\begin{itemize}
    \item  Random, where $\mathcal{U}^{\textrm{rank}}$ is obtained by randomly ordering users.
    \item  InDegree (answer count) \cite{zhang2007expertise}, where $\mathcal{U}^{\textrm{rank}}$ is obtained considering that a user who has given the number of positively scored answers has higher authority.
    \item Z-score \cite{zhang2007expertise}, where $\mathcal{U}^{\textrm{rank}}$ is deduced from considering that a user who answers many questions is an authority, while those who ask more question have less authority.
    \item  Tag-based matrix factorization (T-MF) ~\cite{yang2014tag}. In this approach user-tag activity matrix is decomposed to learn user and tag latent features and subsequently user expertise on a given tag. In other words, the communities that we consider here are unitary. The number of latent features is set to 10, as was done in  previous studies~\cite{yang2014tag}.

    \item Topic Community-based Question Routing (TC-QR), i.e., our approach without the temporal discounting. In this approach user-topic activity matrix is decomposed. Similar to our proposed approach, the number of latent features is set to 10.
\end{itemize}

In our approach, the number of latent features is set as 10, and the regularization parameter $\lambda$ is set as 0.01. The experiments are run on Macbook Pro (CPU: 3.5 GHz Intel Core i7, Memory: 16GB). Matrix factorization is implemented using \textsc{Surprise} library~\cite{hug2020surprise}, community detection is implemented using \textsc{CD-Lib} library \cite{rossetti2019cdlib}. It took less than a minute to run each iteration for both the above steps.

\section{Results}
\subsection{Performance}
Table~\ref{tab_perf} summarises the performance of our proposed model in the datasets given the three performance metrics. 

\begin{table}[!ht]
\centering
\caption{Performance of proposed approach and baseline models.} \label{tab_perf}

\begin{tabular}{|l|l|l|l|l|l|l|l|l|l|}
\hline
{\bf } & \multicolumn{3}{|l|}{\bf Super User} &
\multicolumn{3}{|l|}{\bf Server Fault} & \multicolumn{3}{|l|}{\bf AskUbuntu}\\ 
\hline
Approach & 
MRR & P@5 & P@10 & 
MRR & P@5 & P@10 &
MRR & P@5 & P@10  \\ \hline

Random\footnotemark[1] & 
0.0001 & 0.0005 & 0.001 & 
0.0001 & 0.0009 & 0.002 &
0.0001 & 0.0006 & 0.001 \\ \hline
 
InDegree & 
0.115 & 0.201 & 0.278 & 
0.153 & 0.157 & 0.226 &
0.050 & 0.067 & 0.135 
\\ \hline
Z-score & 
0.140 & 0.201 & 0.276 & 
0.154 & 0.181 & 0.224 &
0.063 & 0.067 & 0.135 
\\ \hline
T-MF & 
0.136 & 0.217 & 0.274 & 
0.099 & 0.162 & 0.274 &
0.078 & 0.102 & 0.183 
\\ \hline
TC-QR & 
0.175 & 0.273 & 0.351 & 
0.188 & 0.285 & 0.358 &
0.127 & 0.206 & 0.280
\\ \hline
TCTE-QR & 
\textbf{0.227} & \textbf{0.324} & \textbf{0.396} &
\textbf{0.255} & \textbf{0.335}& \textbf{0.416} &
\textbf{0.153} & \textbf{0.196} & \textbf{0.328} 
\\ \hline
\end{tabular}

\textit{\footnotetext[1]{The values reported are average scores of 1000 random rankings.}}

\end{table}

The results show that the proposed model significantly outperforms all alternatives. First, the proposed approach without considering temporal dynamics (TC-QR) shows a $\approx40\%$ performance improvement on average across datasets when compared with the T-MF method. Further, with the addition of temporal discounting (TCTE-QR) the performance further improves to $\approx95\%$ over T-MF method.
Our model also outperforms all the baseline models when compared on P$@r$, obtaining an improvement of $\approx80\%$ and $\approx60\%$ for P$@5$ and P$@10$ over T-MF method respectively.

\subsection{Topic-Communities underlying in the TAG-Network}

\subsubsection{Density of User Activity Matrix} 

For the data we used, we found that the sparsity of the user-tag matrix can be high ($\approx99.7\%$). However, since a lot of tags are correlated, clustering tags into topic communities improve the information on users’ expertise as well the density of the user activity matrix. Table~\ref{Sparsity_Eval} shows the result of the user-topic activity matrix density for our approach (TCTE-QR) compared with the user-tag activity matrix in the tag-based approach (T-MF). The density of the matrix increased $\approx45$ times.

\begin{table}[!ht]
\centering
\caption{
 Density of the user activity matrix, defined as the fraction of non-empty entries of $\mathbf{S}$ when the activity is based on tags (first column), and on topic communities. }
\label{Sparsity_Eval}
\begin{tabular}{|l|l|l|}
\hline
CQA & User-Tag  & User-Topic \\ \hline  
Super User & 0.3\% & 15.8\%  \\ \hline 
Ask Ubuntu & 0.4\% & 17.9\% \\ \hline
Server Fault & 0.4\%  & 18.7\% \\\hline 
\end{tabular}
\end{table}

\subsubsection{Significance of the modular structure in the tag network}
To evaluate the effectiveness of the underlying topic communities in $\mathcal{G}$, we compared the performance of the proposed approach based on topics inferred using the original network, and topics inferred using a randomly generated graph.

To generated a random graph, we defined a function that randomly rewires all the edges while preserving the original graph's degree distribution. This is done by choosing two arbitrary edges e.g.\ (a,b) and (c,d), and substituting them with (a,d) and (c,b) if they do not already exist in the original graph. We repeat this until all the edges are swapped to generate a complete random graph (average of $\approx7500$ edge swaps in the three datasets). For the given randomised graph, we perform community detection and compare the performance of TCTE-QR using topic communities inferred from the original network and the randomised network. We measured the performance metrics on ten independent test datasets---alternative quarter datasets starting from Q3-2015 to Q1-2020. For each of the test dataset, the previous two years of data is used as a training dataset. For example, for the Q1-2018 test dataset, we used all the data from $1^{\textrm{st}}$ January 2016 until $31^{\textrm{st}}$ December 2017 as training data. We compared the performance metrics using the Wilcoxon signed-rank test, a widely used statistical test for comparing the performance of recommendation systems~\cite{shani2011evaluating}. Results, shown in Table~\ref{perf_RandomG} suggest that the advance of our technique is due to accounting for the modules within the data resulting in effective clustering of tags in topic communities.

Further, to analyze the effectiveness of clustering tags into topics without temporal discounting, we looked into the performance of TC-QR when topics are obtained from the original network and those inferred from a random graph. Results, shown in Table~\ref{perf_RandomG_TCQR}, suggest that the performance of the TC-QR on topics inferred from the random graph is comparable to T-MF where the tags were treated independently. Further, while the density of the user-topic matrix increases regardless of how nodes are clustered together, the improvement in performance is significantly higher in TC-QR (original graph). These results suggest that reducing sparsity for MF is most effective only when community structure exists and is accounted for in the tag network.

\begin{table}[!ht]
\centering

\caption{Community structure of the tag network $\mathcal{G}$ and performance of TCTE-QR when communities are obtained from considering modularity maximisation (``Original'') and when communities are created from random graph (``Random''). \label{perf_RandomG}}

\begin{tabular}{|p{2cm}|l|l|l|l|l|l|}
\hline
{\bf } & \multicolumn{2}{|l|}{\bf Super User} 
& \multicolumn{2}{|l|}{\bf Server Fault} 
& \multicolumn{2}{|l|}{\bf AskUbuntu} 
\\ \hline

CD Algo. & 
Original & Random & 
Original & Random & 
Original & Random 
\\ \hline

Modularity & 
 0.424 & 0.209 & 0.450 & 0.226 & 0.369 & 0.165 
\\ \hline

User-Topic matrix density & 
 17.6\% & 21.2\% & 13.2\% & 21.3\% & 26.4\% & 24.6\% 
\\ \hline

MRR & 
 0.185 $\bigtriangleup$ & 0.145 & 
 0.174$\bigtriangleup$ & 0.133 & 
 0.147$\bigtriangleup$ & 0.128 
 \\ \hline
P@5 & 
 0.260$\bigtriangleup$ & 0.208 & 
 0.250$\bigtriangleup$ & 0.190 & 
 0.229$\bigtriangleup$ & 0.170
 \\ \hline
P@10 & 
 0.345$\bigtriangleup$ & 0.293 & 
 0.348$\bigtriangleup$ & 0.273 & 
 0.338$\bigtriangleup$ & 0.269 
\\ \hline
\end{tabular}
\end{table}
The table reports average performance measures on the ten test datasets. $\bigtriangleup$ indicate significant improvements over the corresponding recommender. Statistical significance is established by paired Wilcoxon test \cite{woolson2007wilcoxon} with $p$-value threshold of $0.01$ in all cases.

\begin{table}[!ht]
\centering
\caption{Average Performance on ten test datasets of proposed approach (without temporal discounting) and baseline models. } \label{perf_RandomG_TCQR}
\begin{tabular}{|p{1.5cm}|l|l|l|l|l|l|l|l|l|}
\hline
{\bf } & \multicolumn{3}{|l|}{\bf Super User} &
\multicolumn{3}{|l|}{\bf Server Fault} & \multicolumn{3}{|l|}{\bf AskUbuntu}\\ 
\hline
Approach & 
MRR & P@5 & P@10 & 
MRR & P@5 & P@10 &
MRR & P@5 & P@10  \\ \hline

InDegree & 
0.116 & 0.163 & 0.219 & 
0.090 & 0.145 & 0.208 &
0.072 & 0.103 & 0.205 
\\ \hline
Z-score & 
0.116 & 0.162 & 0.218 & 
0.092 & 0.142 & 0.212 &
0.079 & 0.118 & 0.204 
\\ \hline
T-MF & 
0.123 & 0.171 & 0.236 & 
0.112 & 0.157 & 0.226 &
0.099 & 0.144 & 0.219 
\\ \hline
TC-QR (random) & 
0.134 & 0.195 & 0.270 & 
0.114 & 0.172 & 0.262 &
0.085 & 0.117 & 0.209
\\ \hline
TC-QR (original) & 
\textbf{0.155} & \textbf{0.227} & \textbf{0.306} &
\textbf{0.152} & \textbf{0.227}& \textbf{0.314} &
\textbf{0.120} & \textbf{0.180} & \textbf{0.276} 
\\ \hline
\end{tabular}
\end{table}

\paragraph{Robustness of the topic community structure}
Results from Table~\ref{perf_RandomG} suggest that the network has a modular structure, and the modularity of the networks decreases by $\approx50\%$ when the original graph is completely randomised. It is important to note that while networks with high modular structure have high modularity, having high modularity does not imply networks have a modular structure\cite{guimera2004modularity}. Thus, to evaluate the topic communities are significant we follow the approach described in \cite{karrer2008robustness} to access the robustness of the community structure. In particular, we examined the stability of the partition recovered against random perturbations of the original graph. This is done in four steps:

First, we find the community assignment $C$ that maximizes the modularity of the original network for a given community detection algorithm (Louvain in our case). Second, we perturb the network at different perturbation levels p i.e. we rewire a proportion of edges in the original graph and connect them randomly. The value of p is varied from 0 to 1 (20 values), where values close to 0 mean only a few edges moved, while values close to 1 mean network becomes completely random. Third, we find the optimal community assignment $C'$ for that perturbed network and finally measure the variation of information (VI\footnote{an information theoretic criterion that measures the amount of information lost and gained in changing from clustering C to clustering C'})\cite{meilua2007comparing} between $C'$ and $C$. We repeat this step ten times for each value of p to derive an average value for the VI. 

We repeat these four steps starting with a null random graph with the same degree sequence as the original graph and comparing the two VI curves thus obtained. The first curve $VI_{org}$ is obtained computing VI between the partition of the original network and the partition of perturbed versions of the original network. The second curve $VI_{random}$ is obtained computing VI between the partition of a null random network and the partition of perturbed versions of such null network. It is expected that VI should be robust to small perturbation because if small changes in the network result in completely different partition, the found communities are not trustworthy. In other words, robustness/stability of partition against perturbation implies the presence of modular structure in the network \cite{karrer2008robustness,carissimo2018validation}.

\begin{figure}
    \centering
    \includegraphics[scale=0.3]{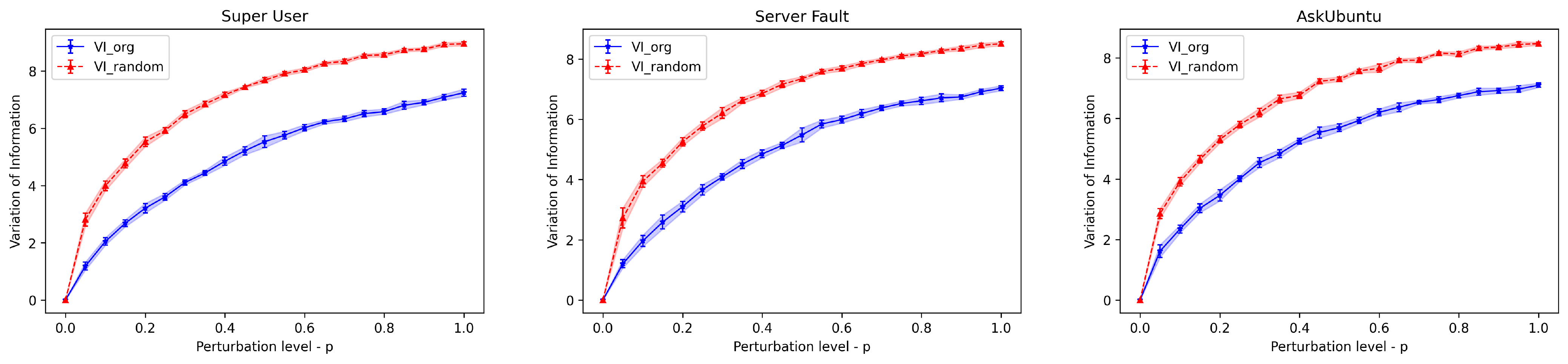}
    \caption{ The variation of information as a function of the perturbation level - p for the three networks (blue), along with equivalent results for the corresponding random graphs(red).}
    \label{perturb}
\end{figure}

 The results in Figure \ref{perturb} show that the community structure discovered in the original network is significantly more robust against perturbations than that of the random graph. For example, while in the case of the original graph the VI value is obtained at perturbation level 0.2 i.e. by rewiring $\approx20\%$ of the edges, for the random graph the same value is obtained by rewiring only $\approx5\%$ of the edges.

\subsubsection{Effectiveness of Community Detection Algorithms}
Although our results relied on Louvain~\cite{blondel2008fast} community detection algorithm, we also considered alternatives. In particular, we compared the results obtained using Louvain algorithm with Leiden~\cite{traag2019louvain} and Greedy modularity~\cite{clauset2004finding} algorithms.

\begin{table}[!ht]
\centering
\caption{Performance based on different Community Detection Algorithms\label{perf_CDAlgo}.}

\begin{tabular}{|*{6}{c|}}
\hline

{\bf CQA}  &{\bf Algorithm}  &{\bf Modularity}  &{\bf MRR}  &{\bf P@5} &{\bf P@10}
\\ \hline

\multirow{3.4}{*}{Super User} & Louvain & 0.397 &  0.227 &  0.324 & 0.396 \\ \cline{2-6}
 & Leiden & 0.427 &  0.219 &  0.314 & 0.393 \\ \cline{2-6}
 & GreedyM & 0.012 &  0.156 &  0.223 & 0.296 \\ \hline

\multirow{3.4}{*} {Server Fault} & Louvain & 0.413 &  0.255 &  0.335 & 0.416 \\ \cline{2-6}
 & Leiden & 0.428 &  0.219 &  0.338 & 0.414 \\ \cline{2-6}
 & GreedyM & 0.013 &  0.179 &  0.215 & 0.321 \\ \hline

\multirow{3.4}{*} {AskUbuntu} & Louvain & 0.334 &  0.153 &  0.196 & 0.328 \\ \cline{2-6}
 & Leiden & 0.354 &  0.152 &  0.221 & 0.327 \\ \cline{2-6}
 & GreedyM & 0.016 &  0.073 &  0.094 & 0.198

\\ \hline
\end{tabular}
\end{table}

Louvain is one of the most widely used community detection algorithms that optimises a quality function such as modularity \cite{newman2004fast} in two phases, first by locally moving nodes to increase the quality function and next by aggregating the network obtained in the previous step. The two phases are repeated until the quality function converges. Leiden algorithm is based on a smart local move algorithm. Greedy-modularity (GreedyM) algorithm works by greedily optimizing the modularity. The algorithm merges two communities at each step that contribute maximum positive value to global modularity. Table~\ref{perf_CDAlgo} shows that the results from Louvain and Leiden algorithm are comparable and were able to find communities with high-quality partitions compared to GreedyM. As a result, tags were clustered more efficiently into topics using the former resulting in the better routing of the questions to the potential expert answerers.

\section{Discussion} 

This paper proposed a topic community-based temporal expertise algorithm for question routing (TCTE-QR) in community-based question answering websites. Our model integrates topic-community detection and accounts for temporal changes in users' activity and interests patterns.  Using community detection we cluster tags, ensuring that correlated items are treated together in matrix factorization. Overall, our method makes use of (i) the modular structure of the underlying tag network, (ii) focuses only on tags instead of all textual information to reduce the computational cost for identifying domain experts, and (iii) the evolution of users' interest and activity patterns. We performed extensive experiments on three real-world datasets which demonstrated significant improvement for question routing over the existing baselines. 
Further, the proposed method improves on the key challenges encountered in the question routing task, namely, the lack of scalability of algorithms, as well as the missing dynamic aspect of the CQA environment \cite{wang2018survey}.

In addition, our work has practical contributions not limited to CQA and can be extended to the recommendation systems domain as a whole. In particular, the data sparsity of the  user-item matrix is a challenge in collaborative filtering and has been addressed by approaches such as data mining, clustering, dimensionality reduction, etc. \cite{najafabadi2016systematic}. Here we argued that this issue can also be solved by leveraging the modular structure of the graph formed by items of interest.

In the future, we intend to address additional issues such as the cold-start problem for improving personalise recommendation in CQA as well extending our approach to other CQA platforms like Quora which has questions from multiple domains. Further, we intend to extend this idea to other datasets in the recommendation domain like music recommendation in Last.fm, movie recommendation on Netflix, product recommendation in Amazon etc.

\paragraph{Acknowledgements:} V.V.\ and N.A.F. acknowledge the support of SoBigData++: European Integrated Infrastructure for Social Mining and Big Data Analytics (Grant agreement ID: 871042).

%%===========================================================================================%%
%% If you are submitting to one of the Nature Portfolio journals, using the eJP submission   %%
%% system, please include the references within the manuscript file itself. You may do this  %%
%% by copying the reference list from your .bbl file, paste it into the main manuscript .tex %%
%% file, and delete the associated \verb+\bibliography+ commands.                            %%
%%===========================================================================================%%

%% if required, the content of .bbl file can be included here once bbl is generated
%%\input sn-article.bbl

%% Default %%
%%\input sn-sample-bib.tex%

%Bibliography
\bibliographystyle{unsrt}  
\bibliography{references}

\end{document}